\documentclass{ws-procs975x65}

\begin{document}



\title{ALIGNED ELECTROMAGNETIC
EXCITATIONS OF THE KERR-SCHILD SOLUTIONS\footnote{Talk at the GT3 session
of the MG11 Meeting, this work is performed in the frame of
collaboration with E.Elizalde, S.R.Hildebrandt and G.Magli.}}

\author{ALEXANDER BURINSKII
}

\address{Gravity Research Group, NSI Russian Academy of Sciences,\\
B. Tulskaya 52, Moscow 115191, Russia, \email{bur@ibrae.ac.ru}}


\def\b{\bar}
\def\d{\partial}
\def\D{\Delta}
\def\cA{{\cal A}}
\def\cD{{\cal D}}
\def\cK{{\cal K}}
\def\cF{{\cal F}}
\def\f{\varphi}
\def\g{\gamma}
\def\G{\Gamma}
\def\l{\lambda}
\def\L{\Lambda}
\def\M{\mathcal{M}}
\def\m{\mu}
\def\n{\nu}
\def\p{\psi}
\def\q{\b q}
\def\r{\rho}
\def\t{\tau}
\def\x{\phi}
\def\X{\~\xi}
\def\~{\tilde}
\def\h{\eta}
\def\bZ{\bar Z}
\def\cY{\bar Y}
\def\bY3{\bar Y_{,3}}
\def\Y3{Y_{,3}}
\def\z{\zeta}
\def\Z{{\b\zeta}}
\def\Y{{\bar Y}}
\def\cZ{{\bar Z}}
\def\`{\dot}
\def\be{\begin{equation}}
\def\ee{\end{equation}}
\def\bea{\begin{eqnarray}}
\def\eea{\end{eqnarray}}
\def\half{\frac{1}{2}}
\def\fn{\footnote}
\def\bh{black hole \ }
\def\cL{{\cal L}}
\def\cH{{\cal H}}
\def\cP{{\cal P}}
\def\cM{{\cal M}}
\def\ol{\overline}
\def\const{{\rm const.\ }}
\def\ik{ik}
\def\mn{{\mu\nu}}
\def\a{\alpha}

\begin{abstract}
 Aligned to the Kerr-Schild geometry electromagnetic excitations
are investigated, and asymptotically exact solutions are obtained
for the low-frequency limit.
\end{abstract}

\bodymatter

{\bf 1.}In this paper we consider the aligned electromagnetic excitations
of the Kerr-Schild geometry, taking into account the back
reaction of the excitations on metric. To our knowledge, it is
the first attempt to get in the Kerr-Schild
formalism a self-consistent solution for the case $\gamma\ne0.$
Electromagnetic field of the exact Kerr-Schild solutions \cite{DKS}
has to be aligned to the Kerr null congruence \cite{BurAxi,BurTwi} which
is generated by tangent vector $k^\m(x).$ Aligned e.m.
excitations on the Kerr background
 were investigated in  \cite{BurAxi,BurTwi,BEHM1}. Contrary to the usual
`quasi-normal' modes, the aligned excitations are compatible with the Kerr
congruence and type D of the metric. On the other hand they have very specific
exhibition in the form of  semi-infinite `axial'
singular lines producing narrow beams which can lead to some new
astrophysical effects like the holes in the horizons and jet
formation \cite{BEHM1}. `Axial' singularities appear also in
the particle aspect of the Kerr-Schild solutions \cite{Multiks}.
\enlargethispage*{6pt}

{\bf 2.} The vector field $k^\m$  is determined by the Kerr
Theorem via a complex function $Y(x),$ $ k_\m dx^\m = P^{-1}(du
+\Y d\z + Y d\Z -Y\Y dv), $
 where  $P$ is a normalizing factor, providing $k_0=1$.
 For the geodesic and shear-free congruences, satisfying to $
Y,_2=Y,_4=0,$ the Einstein-Maxwell field equations were integrated
out in \cite{DKS} in a general form  and reduced to the  system of
equations for electromagnetic field

$ A,_2 - 2 Z^{-1} \cZ Y,_3 A = 0 , \quad A,_4=0 ,$

$ \cD A+ \cZ ^{-1} \gamma ,_2 - Z^{-1} Y,_3 \gamma =0, \quad
\gamma,_4=0 ,$ where $ \cD=\d _3 - Z^{-1} Y,_3 \d_1 - \cZ ^{-1} \Y
,_3 \d_2 $

and for gravitational field, which will be discussed bellow.

{\bf Electromagnetic Sector,} was discussed in
\cite{BurAxi,BurTwi,BEHM1}.  The first equation has the general
solution $ A= \psi/P^2 , $ where $\psi,_2 =\psi,_4=0 .$ Therefore
$\psi$ has to be a holomorphic function of variable $Y$, since
$Y,_2=Y,_4=0.$
 Function $Y$ is a projective (complex) angular
coordinate $Y\in CP^1 =S^2,$ $ Y=e^{i\phi}\tan \frac \theta 2 .$
 A holomorphic
function may be represented as an infinite Laurent series $\psi(Y)
= \sum _{n=-\infty}^{\infty} Y^n.$ If the function $Y\in S^2$ is
not constant, it has to contain at least one pole which may also
be at $Y=\infty$ (or $\theta =\pi$). So, for exclusion of the
Kerr-Newman solution having $Y=e=const. ,$ we has to consider
solutions $ \psi (Y) = \sum _i \frac {q_i} {Y-Y_i} $ which are
singular at angular directions $Y_i =e^{i\phi_i}\tan \frac
{\theta_i} 2 ,$  and represent a narrow beams  in there angular
directions. Note, that for $q_i=const.$ these solutions are exact
self-consistent solutions of the full system of Kerr-Schild
equations.

A wave excitation propagating in the direction $Y_i $ will be
described by the function $ \psi_i(Y,\t) =q(\t)\exp\{i\omega
\t\}\frac 1 {Y-Y_i} .$
where $\t$ is a  retarded time. For the rotating Kerr source the
retarded time is complex\fn{The Kerr solution is described by a
complex `point-like' source propagating along a complex world line
\cite{Bur-nst}. There are different `left' and `right' complex
conjugate world lines and corresponding `left' and `right'
retarded times $\t_L$ and $\t_R.$} In the nonstationary case,
 solution for $A$ has the only difference that the
 function $\psi$ acquires extra dependence from the `left' retarded
 time $\t _L.$ In the rest frame the
 function $P$ has the form $ P=2^{-1/2}(1+Y\bar Y).$
The real operator $\cD$ acts on the real slice as follows $ \cD Y
= \cD \bar Y = 0, $ and $\cD P =0.$ The explicit form of the
retarded time  is $\t _L= t- r + ia \cos \theta .$ Since $\cos
\theta= \frac {1-Y\Y} {1+Y\Y},$ we have $\cD \cos \theta =0,$ ,
and  $ \cD \t =\cD \rho =\frac 1 P .$

 The second e.m. equation takes the form
$ \dot A =-(\gamma P),_{\bar Y} .$ Integration yields
\be \gamma =
 \frac{2^{1/2}\dot \psi} {P^2 Y} +  \phi (Y,\t)/P ,
\label{12}\ee
where we neglected recoil and $\phi$ is an arbitrary analytic
function of $Y$ and $\t$.

{\bf 3.} Gravitational sector is:
\bea M,_2 - 3 Z^{-1} \cZ Y,_3M = A\bar\gamma \cZ , \label{5}  \\
\cD M = \frac 12 \gamma\bar\gamma  , \quad M,_4=0 . \label{6}\eea
Solutions of this system were given in \cite{DKS} only for
stationary case, corresponding to  $\gamma=0$. We assume that
the energy of electromagnetic wave excitation is much lower
then the mass of rotating object $m,$ and does not affect on
the motion of the center of mass of the solution.
However, influence of the electromagnetic field on the metric occurs
also via the function $ H= \frac {mr-\psi\bar\psi /2} {r^2 + a^2 \cos^2
\theta},$ in the K-S metric form $ g_{\m\n} = \h_{\m\n} + 2 H k_{\m}
k_{\n},$ where $ \h_\mn $ is the metric of auxiliary Minkowski
space-time. This is a more thin effect, leading to a deformation
of the metric tensor around rotating black hole by electromagnetic
excitations. The poles in function $\psi$ which cause the `axial'
singular electromagnetic beams deform strongly the function $H$.

 The equation (\ref{5}) acquires the form
$(MP^3),_2 = A\bar \gamma \bar Z .$  The equation (\ref{6})  takes
the form $ \dot m =\frac 12 P^4 \gamma \bar \gamma .$ It is known
\cite{KraSte,Bur-nst}, that it determines the loss of mass by
radiation. The right sides of  (\ref{5}) and (\ref{6}) will be
small for the small (low-frequency) aligned wave excitations,
since the functions $\dot \psi $ and $\gamma$ will be
of order $\sim i\omega \psi .$ In this sense the aligned
excitations will be asymptotically exact solutions in the
low-frequency limit. However, since $\psi$ contains the singular
poles in $Y$, the limit $\gamma \to 0$ is not uniform one, and an
extra trick is necessary - a {\it regularization}.
Such a regularization may be performed by
  the free function $\phi(Y,\t)$ in (\ref{12}).
The function $\gamma$ is represented as a sum of simple poles
$\sum_i \frac {a_i(Y_i,\Y_i, \t) - b_i(Y_i, \t)P_i}{P_i^2(Y-Y_i)},$
where the coefficients $a_i$ are determined by function $\dot \psi,$
and coefficients $b_i$ are chosen from free function $\phi$ to provide
cancelling of the poles. It allows us to perform regularization of the
most of poles in $\gamma$.
If all the poles in the function $\gamma$ will be cancelled, the result of
 integration will be a stochastic radiation which will reduce to zero
for weak excitations, and solutions of (\ref{5}) and (\ref{6}) will be
asymptotically exact.
However, the pole at $Y=\infty$ can not be regularized
by this method and demands especial treatment.

{\bf4.} Structure of the solutions near the beams (pp-waves) is
discussed in \cite{BurAxi,BEHM1}. It was shown that such beams
pierce the horizons forming the tube-like holes connecting internal
and external regions. So the classical structure of black hole
turns out to be destroyed.

Our solution turns out to be exact in the asymptotic limit
$\gamma \to 0 ,$ which corresponds to the weak and slowly changed
electromagnetic field. In particular, it shall tend to exact
one for a black hole immersed into the zero point field of
virtual photons.  In this case we have a sum of excitations in diverse
directions $\psi(Y,\t) = \sum_i \frac {q_i(\t)} {Y-Y_i}$ which
leads to a flow and migration of many singular beams leading to an
instantaneous appearance and disappearance of the holes in horizon,
as it is shown on fig.1. One can assume that it may be a
mechanism of BH evaporation.

\begin{figure}[ht]
\centerline{\epsfig{figure=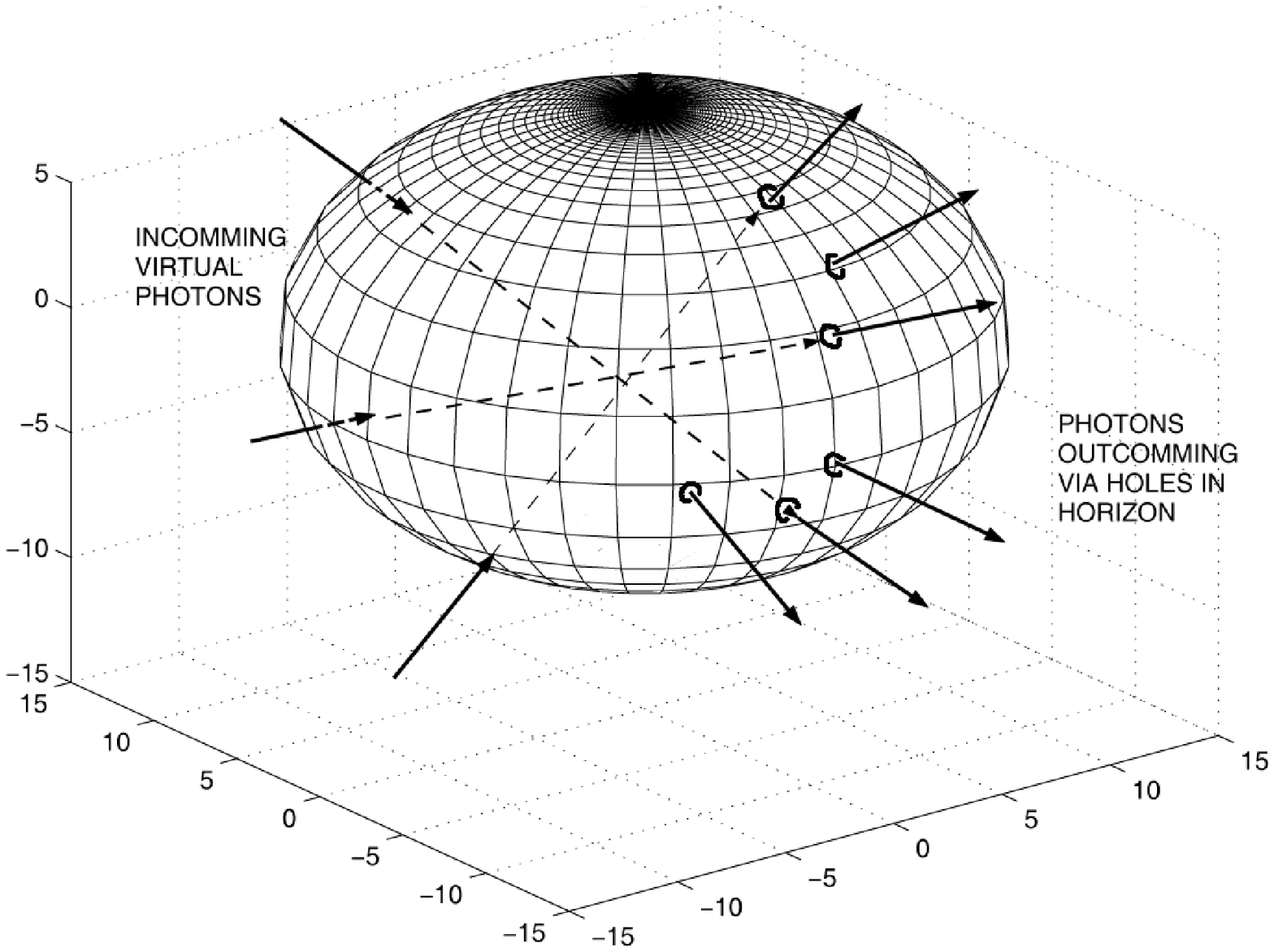,height=4cm,width=7cm}}
\caption{The vacuum flow
 of virtual photons pierces the  black hole horizon.} \end{figure}

Note, that this picture is reminiscent of the haired black hole
which was suggested by the approach from the loop quantum
gravity, where singular hairs were formed from the horizon
contrary to the appearance of the holes in horizon \cite{ABK}.

\vfill

\end{document}